\documentclass[12pt]{iopart}
\usepackage{iopams}
\usepackage[pdftex]{hyperref}
\begin{document}

\eqnobysec
\newcommand{\be}{\begin{equation}}
\newcommand{\ee}{\end{equation}}

\def\link{\prec\!\!*\,}

\begin{flushright}
Imperial/TP/08/SJ/1
\end{flushright}

\title{Particle propagators on discrete spacetime}
\author{Steven Johnston}
\address{Blackett Laboratory, Imperial College London, Prince Consort Road, London, SW7~2AZ, UK.}
\ead{steven.johnston02@imperial.ac.uk}
\date{\today}

\begin{abstract}
A quantum mechanical description of particle propagation on the discrete spacetime of a causal set is presented.
The model involves a discrete path integral in which trajectories within the causal set are summed over to
obtain a particle propagator. The sum-over-trajectories is achieved by a matrix geometric series. For causal
sets generated by sprinkling points into 1+1 and 3+1 dimensional Minkowski spacetime the propagator calculated
on the causal set is shown to agree, in a suitable sense, with the causal retarded propagator for the
Klein-Gordon equation. The particle propagator described here is a step towards quantum field theory on causal
set spacetime.
\end{abstract}


\section{Introduction}

Causal set theory is an approach to quantum gravity in which spacetime is fundamentally discrete. Spacetime
events are represented by elements of a causal set---a locally finite, partially ordered set in which the
partial order represents the causal relationships between events in spacetime. The reader is directed to \cite{Causet1, Causet2} for detailed introductions, motivations and further references.

Modelling spacetime as a causal set is a radical departure from the usual model of spacetime as a Lorentzian
manifold. As such we expect the concepts used to describe physics on a Lorentzian manifold will not be directly
applicable to describing physics on a causal set. In particular the differential equations of motion used in the
continuum are clearly not appropriate (at least in an unmodified form) if spacetime is modelled as a discrete
structure. The discrete path integrals presented here (see Section \ref{Sec:PathIntegrals}) are an example of
applying continuum ideas to a causal set in a natural way.

The appropriate way to model matter on a causal set is an important physical question which, if answered, could
provide testable predictions from the theory. The best description of matter is currently quantum field theory
and we expect that each reader will have his or her own mental picture for interpreting quantum field theory---a
picture based on fields, waves, particles or some combination of these. In this paper we take the viewpoint that
matter is modelled as point particles (rather than fields) propagating according to quantum mechanical laws.
This is in keeping with the propagator approach to quantum field theory pioneered by Feynman (See
\cite{BjorkenDrell}). A particle description rather than a field description is also, perhaps, more in keeping
with the ``discrete'' causal set philosophy.

In Section \ref{Sec:PathIntegrals} we perform quantum mechanical path integrals (or rather path \emph{sums}) on
a causal set. Due to the discrete nature of a causal set these can be defined as actual sums over all particle
trajectories (which move forwards in time) leading from one spacetime event to another. A particle propagator is defined
by summing amplitudes assigned to particle trajectories which move forwards in time. We show that, for causal
sets generated by sprinklings into 1+1 and 3+1 dimensional Minkowski spacetime, this propagator gives (in a
suitable sense) the Klein-Gordon retarded causal propagator from quantum field theory. This describes the
propagation of free scalar particles provided there is no actual pair production in the vacuum\cite{Schwinger}.

Applying the path integral formalism to discrete spacetime has been considered before. Jacobson \cite{Jacobson}
and Gudder \cite{Gudder} have separately considered path integrals on a hypercubic lattice and similar path
integral ideas for causal set spacetime have been considered by Meyer \cite{MeyerTalk} and Foster and Jacobson
\cite{FosterJacobson}. Previous approaches to describe matter on discrete spacetime have often relied on
finite-differencing the continuum equations (see, for example, \cite{Yamamoto}). If we are developing a
fundamentally discrete theory, however, the continuum description should \emph{emerge} from the underlying
theory. The path integrals presented here are an example of an underlying discrete theory leading to a continuum
propagator.

\section{Definitions} \label{Sec:Definitions}

A \emph{causal set} (or \emph{causet}) is a locally finite partially ordered set. This means it is a pair
$(C,\preceq)$ with a set $C$ and a partial order relation $\preceq$ defined on $C$ that is
\begin{itemize}
    \item Reflexive: $x \preceq x$,
    \item Antisymmetric: $x \preceq y \preceq x$ implies $x = y$,
    \item Transitive: $x \preceq y \preceq z$ implies $x \preceq z$,
    \item Locally finite: $\left| \{z \in C | x \preceq z \preceq y\}\right| < \infty$
    for all $x, y \in C$,
\end{itemize}
for all $x, y, z\in C$. Here $\left| A \right|$ denotes the cardinality of a set $A$. We write $x \prec y$ to
mean $x \preceq y$ and $x \neq y$.

The set $C$ represents the set of spacetime events and the partial order $\preceq$ represents the causal order between pairs of events. If $x \preceq y$ we say ``$x$ precedes $y$'' or ``$x$
is to the causal past of $y$''.

A \emph{chain} in a causal set $C$ is a subset in which any two elements are related. If the chain is finite it
is a sequence of distinct elements $v_{i_0}, \ldots, v_{i_n}$ each of which precedes the next element in the
sequence: $v_{i_0} \prec v_{i_1} \prec v_{i_2} \prec \ldots \prec v_{i_n}$. The length of the chain is $n$.

A \emph{link} (or \emph{covering relation}) between $u, v \in C$ is a relation $u \prec v$ such that there
exists no $w \in C$ with $u \prec w \prec v$. We say $u$ and $v$ are \emph{nearest neighbours} (or $v$
\emph{covers} $u$) and write $u \link v$.

A \emph{path} is a subset $P \subset C$ which is a maximal (or saturated) chain. This means $P$ is a chain with
no element $w \in C - P$ such that $u \prec w \prec v$ for some $u, v \in P$. If the path is finite it is a
sequence of distinct elements $v_{i_0}, \ldots, v_{i_n}$ each of which is linked to the next element in the
sequence: $v_{i_0}\link v_{i_1} \link v_{i_2} \link \ldots \link v_{i_n}$. The length of the path is $n$, the
number of links.

\subsection{Adjacency matrices} \label{Sec:AdjacencyMatrices}

A useful way to represent a finite causal set is by its \emph{adjacency matrix}. We now define two types of
adjacency matrices that will be used to compute the path integrals in Section \ref{Sec:PathIntegrals}. We begin
by looking at finite causal sets as this allows us to work with finite square matrices. In Section
\ref{Sec:GeneralCausalSets} we extend the work to general causal sets.

Before defining the adjacency matrices we must label the elements in our causal set. For a finite causal set $C$
with $p$ elements we choose a labelling $v_1,\ldots,v_p$ for the elements of $C$. There are then two adjacency
matrices of interest: $A_C$ and $A_R$. These are $p\times p$ matrices defined by \be (A_C)_{ij} := \left\{
\begin{array}{ll} 1 & \textrm{ if } v_i \prec v_j \\ 0 & \textrm{otherwise,} \end{array} \right. \ee \be
(A_R)_{ij} := \left\{
\begin{array}{ll} 1 & \textrm{ if } v_i \link v_j \\ 0 & \textrm{otherwise.} \end{array} \right. \ee These are
both zero on the main diagonal and, from the definition of a causal set, the labelling can always be chosen to
ensure they are both strictly upper triangular matrices (in which case the labelling is called a \emph{natural
labelling}). Within graph theory, the matrix $A_C$ is called the transitively closed adjacency matrix and $A_R$
is called the transitively reduced adjacency matrix. Sorkin \cite{CausetLocality} has termed $A_C$ the
\emph{causal matrix} and $A_R$ the \emph{link matrix}.

Powers of these matrices have the following useful properties\cite{EnumerativeCombinatorics}: \be
\left(A_C^n\right)_{ij} = \textrm{The number of chains of length $n$ from $v_i$ to $v_j$,}\ee \be
\left(A_R^n\right)_{ij} = \textrm{The number of paths of length $n$ from $v_i$ to $v_j$.}\ee

\subsection{Sprinklings} \label{Sec:Sprinklings}

There exist causal sets which bear little resemblance to the smooth Lorentzian manifolds of general relativity.
It is hoped that in a full theory of causal set dynamics the causal sets which are ``manifold-like'' on large
scales will emerge dynamically from the theory. For the time being, however, if we are interested in comparing
the results of calculations performed on a causal set with calculations on a Lorentzian manifold we must ensure
by hand that the causal set is like the manifold.

We do this by performing a \emph{sprinkling} (see \cite{Sprinkling} for more details) into a $d$-dimensional
Lorentzian manifold $(M,g)$. To perform a sprinkling points are placed at random within $M$ using a Poisson
process with sprinkling density $\rho$. This means that the points are randomly distributed throughout $M$ with
$n$ points being sprinkled into a region of $d$-volume $V$ with probability \be \label{eq:PoissonDef}
\textrm{Prob($n$ points in region of $d$-volume $V$)} = \frac{(\rho V)^n}{n!}e^{-\rho V}. \ee This ensures the
expected number of points in a region of $d$-volume $V$ is equal to $\rho V$. The sprinkling generates a causal
set whose elements are the sprinkled points. The causal set's partial order relation is ``read off'' from the
manifold's causal relation restricted to the sprinkled points. If the Lorentzian manifold satisfies a strong
enough causality condition (e.g. if it's strongly causal) then we expect the manifold to well-approximate the
sprinkled causal set.

We note that setting $n=0$ in \eref{eq:PoissonDef} gives the probability that two points $x \prec y$ in $M$ have no
sprinkled points causally between them as $e^{-\rho V(x-y)}$. Here $V(x-y)$ is the $d$-volume of the causal
interval in $M$ between $x$ and $y$.

\section{Path integrals on causal sets} \label{Sec:PathIntegrals}

A classical model for particle propagation is the swerves model described in \cite{CausetSwerves}. As
acknowledged in that paper, however, a fundamental description should incorporate the principles of quantum
mechanics.

In the continuum the propagation of particles is described by a propagator $K(x,y)$ which can be interpreted as
the quantum mechanical amplitude (or just amplitude) for a particle to travel between two spacetime points $x$
and $y$ (see, for example, \cite[chapter 6]{BjorkenDrell}). The path integral approach to quantum mechanics
\cite{FeynmanHibbs, Feynman2} allows the propagator to be calculated by assigning amplitudes to each possible
particle trajectory from $x$ to $y$. Summing these amplitudes over all possible trajectories from $x$ to $y$
gives the propagator $K(x,y)$. As we shall show, in the discrete spacetime of a causal set this sum over trajectories can be simply performed.

To define path integrals on a causal set we have to make two choices: which trajectories to sum over and what
amplitudes to assign to each trajectory. In this paper we'll consider two types of trajectories within the
causal set: either all possible chains between two elements or all possible paths between two elements. The
amplitude assigned to each trajectory is then defined in terms of two basic amplitudes: $a$ and $b$.

The constant $a$ is the amplitude for the particle to `hop' once along the trajectory from one element to the next.
The constant $b$ is the amplitude for the particle to `stop' at one element of the trajectory (the initial and
final elements are not regarded as stops). The amplitude for the whole trajectory is then the product of the
amplitudes for each hop and each stop. For a chain or path of length $n$ (so there are $n$ hops and $n-1$
intermediate stops) this compound amplitude is therefore $a^n b^{n-1}$.

In Section \ref{sec:MinkowskiPropagators} we identify values of $a$ and $b$ which, for causal sets generated by
sprinkling into 1+1 and 3+1 dimensional Minkowski spacetime, give (in a sense to be defined) the causal
retarded Klein-Gordon propagator. These values (equations \eref{eq:1+1Amplitudes} and \eref{eq:3+1Amplitudes})
depend on the mass of the particle and on the spacetime volume assigned to each causal set point.

As in Section \ref{Sec:AdjacencyMatrices} we begin by working with finite causal sets. How the ideas extend to
arbitrary causal sets is presented in Section \ref{Sec:GeneralCausalSets}. For a causal set with $p$ elements we
define a $p \times p$ matrix $\Phi$.

If we sum over chains we define \be \Phi:= a A_C. \ee If we sum over paths we define \be \Phi:= a A_R. \ee The
total amplitude to go from $v_i$ to $v_j$ along a trajectory of any length is then $K_{ij}$ where $K$ is the $p
\times p$ matrix \be \label{eq:CausetPropagator} K:= I + \Phi + b\Phi^2 + b^2\Phi^3 + \ldots = I +
\sum_{n=1}^\infty b^{n-1} \Phi^n.\ee Here matrix multiplication is used and $I$ is the $p\times p$ identity
matrix. We have included the $I$ matrix to ensure $K_{ii} = 1$ for all $i=1,\ldots,p$ although this is just
convention.

Each term in this sum is the contribution to the total amplitude from chains or paths of a particular length. As
an example, \be b(\Phi^2)_{ij} = \sum_{k=1}^p b \Phi_{ik} \Phi_{kj}, \ee is the sum (over all intermediate
positions $v_k$) of the amplitudes for all length two trajectories $v_i$ to $v_k$ to $v_j$.

Since the causal set is finite and the trajectories move only forwards in time the sum in
\eref{eq:CausetPropagator} terminates and we have \be K = I + \Phi(I - b \Phi)^{-1}. \ee This matrix inverse is
simple to perform if the causal set has been labelled to ensure $\Phi$ is strictly upper triangular. In this
case the upper triangular matrix $I - b\Phi$ can be inverted using elementary row operations.

The question we now face is whether there are values of $a$ and $b$ such that $K_{ij}$, for a causal set
generated by a sprinkling into Minkowski spacetime, is approximately equal to the continuum propagator in
Minkowski spacetime. We address this in the next section.

\subsection{Expected values}

Causal sets generated by sprinklings into a Lorentzian manifold will not be identical because the particular
points that are sprinkled are chosen randomly. The value of the propagator calculated from one sprinkled causal
set therefore depends in detail on the particular causal set that is generated. This randomness in the
sprinkling process means we are interested in the \emph{expected value} (and variance) for the propagator calculated
by averaging over all possible sprinkled causal sets.

We shall look at sprinklings into $d$-dimensional Minkowski spacetime. Firstly we shall fix two points $x$ and
$y$ in Minkowski spacetime. We then sprinkle a causal set. There is zero probability that the sprinkled causal
set will contain $x$ and $y$ so we then add $x$ and $y$ to the sprinkled causal set. In this manner, averaging
over all sprinkled causal sets (with a fixed sprinkling density $\rho$), we shall calculate the expected number
of chains and paths from $x$ to $y$ and the expected value of the propagator between $x$ and $y$.

\subsubsection{Summing over chains}

To calculate the expected number of chains between two points $x$ and $y$ in $d$-dimensional Minkowski spacetime
we define \be \nu(x-y) := \left\{ \begin{array}{ll} 1 & \textrm{ if } x \preceq y \\ 0 & \textrm{otherwise.}
\end{array} \right. \ee
Translation invariance of Minkowski spacetime ensures $\nu$ is only a function of the separation $x-y$. The
expected number of chains of length one from $x$ to $y$ is given by \be C_1(x-y) = \nu(x-y). \ee The expected
number of chains $x \prec z_1 \prec \ldots \prec z_{n-1} \prec y$ of length $n > 1$ is given by\cite{MeyerTalk}
\be \fl C_n (x-y) := \rho^{n-1}\int \cdots \int \rmd^d z_1 \cdots \rmd^dz_{n-1} \nu(x-z_1) \nu(z_1 - z_2) \cdots
\nu(z_{n-1}-y). \ee The expected value for a propagator which sums amplitudes assigned to chains is thus \be
\label{eq:ChainInfiniteSum} K_\mathrm{C}(x-y) := \sum_{n=1}^\infty a^n b^{n-1}  C_n (x-y). \ee This satisfies the
integral equation \be \label{eq:ChainSumExpectedValue} K_\mathrm{C}(x-y) = a \nu(x-y) + ab \rho \int \rmd^dz\, \nu(x-z)
K_\mathrm{C}(z-y). \ee

\subsubsection{Summing over paths}

To calculate the expected number of paths between two sprinkled points $x$ and $y$ in $d$-dimensional Minkowski
spacetime we define \be \mu_\rho(x-y) := \left\{
\begin{array}{ll} e^{- \rho V(x-y)} & \textrm{ if } x \preceq y \\ 0 & \textrm{otherwise.} \end{array} \right.
\ee where $V(x-y)$ is the $d$-dimensional Minkowski spacetime volume of the causal interval between $x$ and $y$.
The expected number of paths of length one from $x$ to $y$ is given by \be P_1(x-y) = \mu_\rho(x-y).\ee The
expected number of paths $x \link z_1 \link \ldots \link z_{n-1} \link y$ of length $n > 1$ is given
by\cite{BombelliThesis} \be \fl P_n (x-y) = \rho^{n-1}\int \cdots \int \rmd^d z_1 \cdots \rmd^dz_{n-1} \mu_\rho(x-z_1)
\mu_\rho(z_1 - z_2) \cdots \mu_\rho(z_{n-1}-y). \ee The expected value for the propagator which sums over paths
is thus \be \label{eq:PathInfiniteSum} K_\mathrm{P}(x-y) := \sum_{n=1}^\infty a^n b^{n-1}  P_n (x-y). \ee This satisfies
the integral equation \be \label{eq:PathSumExpectedValue} K_\mathrm{P}(x-y) = a \mu_\rho(x-y) + ab\rho \int \rmd^dz\,
\mu_\rho(x-z) K_\mathrm{P}(z-y). \ee

\subsection{Minkowski spacetime propagators} \label{sec:MinkowskiPropagators}
In $d$-dimensional Minkowski spacetime the causal relations are particularly simple. For $x = (x_0,\vec{x}),\,y
= (y_0,\vec{y})$ we have: \be x \preceq y \iff x_0 \leq y_0 \quad \textrm{and} \quad (x_0-y_0)^2 \geq
(\vec{x}-\vec{y})^2.\ee

The scalar propagators in $d$-dimensional Minkowski spacetime quantum field theory are Green's functions of the
Klein-Gordon equation: \be \label{eq:KleinGordon} (\Box + m^2)K^{(d)}_m(x-y) = \delta^d(x-y).\ee Here $m$ is the
mass of the particle, $\delta^d$ is the $d$-dimensional Dirac delta function and we choose units with $\hbar = c
=1$. The d'Alembertian is given by \be \Box:= \frac{\partial^2}{\partial {x_0}^2} - \frac{\partial^2}{\partial
{x_1}^2} - \frac{\partial^2}{\partial {x_2}^2} - \ldots - \frac{\partial^2}{\partial {x_{d-1}}^2}. \ee To obtain
the propagator $K^{(d)}_m(x-y)$ explicitly we define the Fourier transform by \be \widetilde{f}(p) := \int \rmd^dx
f(x) e^{i p x}, \qquad f(x) = \frac{1}{(2 \pi)^d}\int \rmd^d p \widetilde{f}(p) e^{-ip x}, \ee where $p x := p_0
x_0 - \vec{p}\cdot \vec{x}$. Using this we can solve equation \eref{eq:KleinGordon} to obtain \be
\widetilde{K}^{(d)}_m(p) := -\frac{1}{p_0^2 - \vec{p}^2 - m^2},\ee so \be K^{(d)}_m(x-y) := -\frac{1}{(2 \pi)^d}
\int \rmd^d p \frac{e^{-ip(x-y)}}{p_0^2 - \vec{p}^2 - m^2}. \ee The mass-dimension of this propagator is
$[K^{(d)}_m] = M^{d-2}$.

The integrand in these expressions contains poles so the propagator in Minkowski spacetime is only uniquely
defined if we specify a contour of integration (or equivalently boundary conditions for the solution of
\eref{eq:KleinGordon}). The \emph{retarded causal propagator} is the Green's function obtained by avoiding the
poles at $p_0 = \pm \sqrt{\vec{p}^2 + m^2}$ by two small semi-circles in the upper-half $p_0$ complex plane.
This is equivalent to \be K^{(d)}_m(x-y) := \lim_{\epsilon \to 0^+} -\frac{1}{(2 \pi)^d} \int \rmd^d p
\frac{e^{-ip(x-y)}}{(p_0+i \epsilon)^2 - \vec{p}^2 - m^2}. \ee This can be calculated explicitly in different
dimensions. We consider the expressions in $d=2$ and $d=4$ dimensions.

In 1+1 dimensional Minkowski spacetime we have: \be \label{eq:2dPosProp} K^{(2)}_m(x-y) = \left\{
\begin{array}{ll} \frac{1}{2}J_0(m \tau_{xy}) & \textrm{ if } x \preceq y \\ 0 & \textrm{otherwise,} \end{array}
\right. \ee \be \label{eq:2dMomProp} \widetilde{K}^{(2)}_m(p) = -\frac{1}{(p_0+i\epsilon)^2 - p_1^2 - m^2}. \ee
Here $\tau_{xy} = \sqrt{(x_0-y_0)^2 - (x_1 - y_1)^2}$ is the proper time from $x$ to $y$ and $J_0$ is the
zeroth-order Bessel function of the first kind.

In 3+1 dimensional Minkowski spacetime\cite{KGSolutions}: \be \label{eq:4dPosProp} K^{(4)}_m(x-y) = \left\{
\begin{array}{ll} \frac{1}{2 \pi} \delta(\tau^2_{xy}) - \frac{m}{4\pi} \frac{J_1(m \tau_{xy})}{\tau_{xy}} &
\textrm{ if } x \preceq y \\ 0 & \textrm{otherwise,} \end{array} \right. \ee \be \label{eq:4dMomProp}
\widetilde{K}^{(4)}_m(p) = -\frac{1}{(p_0+i\epsilon)^2 - \vec{p}^2 - m^2}. \ee Here $\tau_{xy} =
\sqrt{(x_0-y_0)^2 - (\vec{x} - \vec{y})^2}$ is the proper time from $x$ to $y$ and $J_1$ is the first-order
Bessel function of the first kind.

We keep the $\epsilon$ terms in the Fourier transformed propagators to ensure the poles are avoided in the
correct way. It is understood that the $\epsilon \to 0^+$ limit is taken at the end of the calculations. It is
these position-space propagators we wish to reproduce on a causal set.

\subsubsection{1+1 dimensional Minkowski spacetime}

It turns out that in 1+1 dimensions the propagator requires us to sum over chains. Fourier transforming
\eref{eq:ChainSumExpectedValue} the integral, being a convolution, becomes a product and we have \be
\widetilde{K}_\mathrm{C}(p) = a \widetilde{\nu}(p) + ab\rho \,\widetilde{\nu}(p) \widetilde{K}_\mathrm{C}(p), \ee or \be
\label{eq:ChainSumMomSpace} \widetilde{K}_\mathrm{C}(p) = \frac{a \widetilde{\nu}(p)}{1 - ab\rho \widetilde{\nu}(p)}. \ee
In 1+1 dimensions the function $\nu(x-y)$ has Fourier transform \be \widetilde{\nu}(p) = -\frac{2}{(p_0 + i
\epsilon)^2 - p_1^2}, \ee since $\nu = 2 K^{(2)}_0$ (and using \eref{eq:2dPosProp} and \eref{eq:2dMomProp} with $m=0$). Substituting into
\eref{eq:ChainSumMomSpace} we have \be \widetilde{K}_\mathrm{C}(p) = \frac{-\frac{2a}{(p_0+i\epsilon)^2 - p_1^2}}{1 +
\frac{2 ab \rho}{(p_0+i\epsilon)^2 - p_1^2}} = -\frac{2a}{(p_0+i\epsilon)^2 - p_1^2 + 2ab\rho}.\ee Equating this
to equation \eref{eq:2dMomProp} we find \be \label{eq:1+1Amplitudes} a = \frac{1}{2}, \qquad b =
-\frac{m^2}{\rho}, \ee are the correct amplitudes.

In 1+1 dimensions $[\rho] = M^2$ so the amplitudes assigned to the chains are dimensionless: $[a] = [b] = [a^n
b^{n-1}] = 1$ (for $n=1,2,\ldots$).

\subsubsection{3 + 1 dimensional Minkowski spacetime}

It turns out that in 3+1 dimensions the propagator requires us to sum over paths. Fourier transforming
\eref{eq:PathSumExpectedValue} we get \be \label{eq:PathSumMomSpace} \widetilde{K}_\mathrm{P}(p) = \frac{a
\widetilde{\mu}_\rho(p)}{1 - ab\rho \widetilde{\mu}_\rho(p)}. \ee In 3+1 dimensional Minkowski spacetime $V(x-y)
:= \frac{\pi}{24} \tau_{xy}^4$ is the volume of the causal interval between $x$ and $y$. We therefore have
\be \mu_\rho(x-y) = \left\{ \begin{array}{ll} e^{-\rho V(x-y)} = e^{-\frac{\pi}{24} \rho \tau_{xy}^4} & \textrm{if } x \preceq y \\
0 & \textrm{otherwise.} \end{array} \right. \ee The function
\be f_\rho(z,c) := \left\{ \begin{array}{ll} \sqrt{\rho} e^{-\pi c \rho z^2} & \textrm{if } z \geq 0 \\
0 & \textrm{if } z < 0\end{array} \right. \ee (with a real constant $c > 0$) satisfies \be \lim_{\rho \to
\infty} f_\rho(z,c) = \frac{1}{2 \sqrt{c}} \delta(z), \ee where $\delta(z)$ is the Dirac delta function. We
therefore see
\be \label{eq:3+1Limit} \fl \lim_{\rho \to \infty} \sqrt{\rho} \,\mu_\rho(x-y) = \lim_{\rho \to \infty} \left\{ \begin{array}{ll} f_\rho(\tau_{xy}^2, \frac{1}{24}) & \textrm{if } x_0 \leq y_0 \\
0 & \textrm{otherwise} \end{array} \right. = \left\{ \begin{array}{ll} \frac{\sqrt{24}}{2} \delta(\tau_{xy}^2) & \textrm{if } x \preceq y \\
0 & \textrm{otherwise.} \end{array} \right. \ee Setting $m=0$ in \eref{eq:4dPosProp} and \eref{eq:4dMomProp}
we see that in 3+1 dimensions we have \be K^{(4)}_0(x-y) = \left\{
\begin{array}{ll} \frac{1}{2 \pi} \delta(\tau^2_{xy})& \textrm{ if } x \preceq y \\ 0 & \textrm{otherwise,}
\end{array} \right. \ee \be \widetilde{K}^{(4)}_0(p) = -\frac{1}{(p_0+i\epsilon)^2 - \vec{p}^2}.\ee
These imply, taking Fourier transforms and using \eref{eq:3+1Limit}, that \be \lim_{\rho \to \infty}
\sqrt{\rho}\; \widetilde{\mu}_\rho(p) = 2 \pi \frac{\sqrt{24}}{2} \widetilde{K}^{(4)}_0(p) = -2 \pi\sqrt{6}
\frac{1}{(p_0+i\epsilon)^2 - \vec{p}^2}.\ee Setting \be a = A \sqrt{\rho}, \qquad b = \frac{B}{\rho}, \ee where
$A$ and $B$ are (possibly dimensionful) constants independent of $\rho$ we substitute into
\eref{eq:PathSumMomSpace} to get \be \widetilde{K}_\mathrm{P}(p) = \frac{A \sqrt{\rho}\widetilde{\mu}_\rho(p)}{1 -
AB\sqrt{\rho}\widetilde{\mu}_\rho(p)}.\ee As $\rho$ tends to infinity this becomes \be \lim_{\rho \to \infty}
\widetilde{K}_\mathrm{P}(p)=\frac{-\frac{A C}{(p_0+i\epsilon)^2 - \vec{p}^2}}{1 + \frac{AB C}{(p_0+i\epsilon)^2 -
\vec{p}^2}}  = -\frac{AC}{(p_0+i\epsilon)^2 - \vec{p}^2 + ABC},\ee where $C := 2\pi\sqrt{6}$.

Equating this with \eref{eq:4dMomProp} we have \be AC = 1, \qquad B = -m^2,\ee so \be \label{eq:3+1Amplitudes}
a = \frac{\sqrt{\rho}}{2 \pi \sqrt{6}}, \qquad b = -\frac{m^2}{\rho},\ee are the correct amplitudes.

In 3+1 dimensions $[\rho] = M^4$ so the amplitudes assigned to the paths have mass-dimension $M^2$: $[a] = M^2$,
$[b] = M^{-2}$, $[a^n b^{n-1}] = M^2$ (for $n=1,2,\ldots)$.

\subsection{Fluctuations}

The basic amplitudes, $a$ and $b$, given in \eref{eq:1+1Amplitudes} and \eref{eq:3+1Amplitudes} ensure that
the expected value of the propagator, averaged over causal sets generated by sprinkling into 1+1 and 3+1
dimensional Minkowski spacetime, equals the causal retarded propagators for the Klein-Gordon equation (with the
infinite density limit being taken in the 3+1 case). The propagator values calculated on any particular
sprinkled causal set will, however, fluctuate randomly away from the expected values. An important question is
how the size of these fluctuations vary as the density of the sprinkling increases. If the fluctuations increase
as the density increases (i.e. in the continuum limit) then the model clearly needs to be changed. An example of
such a situation is given in \cite{CausetLocality} where a causal set approximation to the d'Alembertian has
growing fluctuations in the large density limit. This is resolved by modifying the d'Alembertian approximation
by performing a sum over points contained within a new non-locality scale of the original point.

One way to investigate the question of how the fluctuations depend on the sprinkling density is through
numerical simulations. These involve generating causal sets by sprinkling points into an interval of Minkowski
spacetime. We then calculate the propagator on the generated causal set and examine the fluctuations for
different density sprinklings. The fluctuations depend on the sprinkling density $\rho$ as well as on the size
of the particle's mass $m$ relative to $\rho$. The simulations discussed here use causal sets generated by David
Rideout's Cactus computer code (briefly described in \cite{RideoutCode}).

For sprinklings into 1+1 dimensional Minkowski spacetime and for $m^2 \ll \rho$ preliminary results suggest the
fluctuations decrease as the density increases.

For sprinklings into 3+1 dimensional Minkowski spacetime the propagator's expected value equals the continuum
propagator only in the infinite density limit. A realistic sprinkling density would be large\footnote{Equivalently we can use a small density but sprinkle into a Minkowski spacetime
interval of large volume.} but finite. Investigating the behaviour of the propagator for large finite
sprinkling densities is difficult to do through numerical simulations. This is because current simulations
cannot produce enough sprinkled points to ensure a large density over a large spacetime volume.

High density sprinklings \emph{can} be achieved by sprinkling a moderate number of points into a small volume.
In this case, however, the behaviour of the propagator is only investigated within such small volumes. In 3+1 dimensional Minkowski spacetime the preliminary results from such simulations suggest, for $m^2 \ll \sqrt{\rho}$, the expected value of the propagator for large densities well-approximates the continuum propagator. In addition we may very preliminarily say the fluctuations away from the continuum propagator are small for large density. These conclusions hold only for the small volumes considered in the simulations and further work is needed to check whether this behaviour holds for large density sprinklings into large spacetime volumes. Of course determining the rate at which the fluctuations decrease remains to be done.

We note here that the constraints on the mass of the particle are easily satisfied for realistic sprinkling
densities. Assuming a Planckian sprinkling density in the 3+1 case we let $\rho$ be the inverse of the Planck
4-volume: $\rho =c^7/(G \hbar)^2$. The heaviest known elementary particle is the top quark with a mass $m =
174.2 \pm 3.3$GeV \cite{PDG}. In Planck units, so $\rho = 1$, this is $m \approx 1.4 \times 10^{-17}$. This
certainly satisfies $m^2/\sqrt{\rho} \approx 2 \times 10^{-34} \ll 1$. If the masses of particles modelled on the causal set are of order the masses of known elementary particles then the causal set propagator (for a Planckian density sprinkling) should well-approximate the continuum propagator.

\section{General causal sets} \label{Sec:GeneralCausalSets}

We now consider general causal sets---including non-finite causal sets as well as those not generated by
sprinkling into Minkowski spacetime.

\subsection{Incidence algebra}

An \emph{interval} (or Alexandroff neighbourhood) in a causal set $C$ is defined to be \be[u,v]:=\{ w \in C |\,
u \preceq w \preceq v\},\ee provided $u \preceq v$.

Since a causal set is locally finite the propagator $K(u,v)$ for any two points $u \preceq v$ can be calculated by
applying the methods of Section \ref{Sec:PathIntegrals} to the finite interval $[u,v]$. There is another way,
however, to view the path integral framework which uses the \emph{incidence algebra} of a causal set.

For a causal set $C$ (not necessarily finite) we denote the set of all intervals by \be\textrm{Int}(C):= \{[u,v]
|\, u, v \in C ,\, u \preceq v \}. \ee

The incidence algebra \cite{EnumerativeCombinatorics} of $C$ over $\mathbb{C}$, denoted
$I(C,\mathbb{C})$, is then the associative algebra of all functions \be f: \textrm{Int}(C) \to \mathbb{C}, \ee
with multiplication defined by \be f * g (x,y):= \sum_{x \preceq z \preceq y} f(x,z)g(z,y).\ee The sum here is
finite (so $f * g$ is defined) because the causal set is locally finite. $I(C,\mathbb{C})$ is an associative
algebra with two-sided identity \be \delta(x,y):= \left\{\begin{array}{ll} 1 & \textrm{ if } x = y
\\ 0 & \textrm{ otherwise. } \end{array}\right. \ee We note that we could use an algebraic field other
than $\mathbb{C}$ in defining the algebra. The adjacency matrices, which were defined only for finite causal
sets, generalize to the algebra elements \be A_C(x,y):= \left\{ \begin{array}{ll} 1 & \textrm{ if } x \prec y \\
0 & \textrm{ otherwise, }  \end{array}\right. \ee \be A_R(x,y):= \left\{ \begin{array}{ll} 1 & \textrm{ if } x
\link y \\ 0 & \textrm{ otherwise. } \end{array}\right. \ee Powers of these functions, under $*$, satisfy: \be
A_C^n (x,y) = \textrm{The number of chains of length $n$ from $x$ to $y$.}\ee \be A_R^n (x,y) = \textrm{The
number of paths of length $n$ from $x$ to $y$.}\ee

The path integral work in Section \ref{Sec:PathIntegrals} can be done using the incidence algebra. Phrasing the
method this way rather than using adjacency matrices allows us to work with infinite causal sets without
restricting to a finite sub-causal-set. For finite causal sets the incidence algebra and adjacency matrix
methods are entirely equivalent.

If we sum over chains we define an element $\Phi$ of $I(C,\mathbb{C})$ by \be \Phi(x,y):= a A_C(x,y).\ee If we
sum over paths we define \be \Phi(x,y):=  a A_R(x,y). \ee We then have, in a manner similar to the finite case,
that \be K := \delta + \Phi + b (\Phi * \Phi) + b^2 (\Phi
* \Phi
* \Phi) + \ldots \ee is the algebra element for the propagator. Here $K(x,y)$ is the quantum mechanical
amplitude that the particle will travel from $x$ to $y$ along a trajectory of any length.

Formally we have \be K = \delta + \Phi * (\delta - b \Phi)^{-1},\ee but it is not immediately clear if an
inverse of $\delta - b \Phi$ exists. Applying Proposition 3.6.2 in \cite{EnumerativeCombinatorics} shows,
however, that $(\delta - b \Phi)^{-1}$ exists so $K$ can be written in this form. The proposition also shows
that $K(x,y)$ depends only on the values of $\Phi$ for intervals contained within $[x,y]$.

It is pleasing to note that the choice of modelling spacetime as a causal set enables us to define its incidence
algebra which can then be used to define path integrals. The choice of causal set spacetime fits naturally with
the rules governing quantum mechanical amplitudes.

\subsection{Non-sprinkled causal sets}

The basic amplitudes for sprinkled causal sets depend on the sprinkling density $\rho$. For causal sets not
generated by a sprinkling, however, we must make sense of the $\rho$ that appears in
$a$ and $b$. To do this we assume that an arbitrary causal set element is assigned a fundamental spacetime
volume $V_0$. When the causal set is generated by a sprinkling into Minkowski spacetime with density $\rho$ we
have $V_0 = 1/\rho$.

If the causal set leads to a macroscopically 1+1 dimensional spacetime we sum over chains and assume $[V_0] =
M^{-2}$. This gives \be a = \frac{1}{2}, \qquad b = -m^2 V_0. \ee

If the causal set leads to a macroscopically 3+1 dimensional spacetime we sum over paths and assume $[V_0] =
M^{-4}$. This gives \be a = \frac{1}{2 \pi \sqrt{6 V_0}}, \qquad b = -m^2 V_0. \ee

\section{Further work}

One of the motivations for discrete spacetime is to regularize the divergences in quantum field theory. The
propagators presented here could be used to develop quantum field theory on a causal set and, in particular, the
graph-based language of Feynman diagrams appears particularly well-suited for this. This paper deals only with
scalar particles and, as required for a realistic quantum field theory, we would hope to model spinor and vector
particles on a causal set---possibly by allowing internal particle states. Another interesting question is
whether the Feynman propagator can be obtained by a path integral on a causal set---possibly by allowing
trajectories which can reverse time-direction\footnote{These paths would be interpreted as particle pair
production\cite{BjorkenDrell}}. These remain open questions.

The perspective used in this paper is that of relativistic quantum mechanics---the motion of matter is described
upon a fixed spacetime background (the causal set). No backreaction has been included in which the structure of
the causal set would depend on the matter distribution upon it. This is one of the principal difficulties in
quantum gravity---to obtain a description of matter and spacetime that is able to describe more than just
quantum matter on a fixed classical spacetime background. This may be achieved in the long sought-after ``sum
over causal sets'' version of the theory.

\ack

The author would like to thank Fay Dowker for helpful support. Also Joe Henson and Leron Borsten for early
discussions, David Reid for providing a useful reference and David Rideout for much help with the numerical
simulations. ENRAGE grant MRTN-CT-2004-005616 was used to fund travel associated with this work.
This work was supported by a STFC studentship.

\end{document}